\documentclass[
    ,final            
  ]
  {aipproc}

\layoutstyle{6x9}

\usepackage{pifont}  

\newcommand{\GeV}{\ensuremath{\mathrm{~Ge\kern -0.1em V}}}
\newcommand{\GeVc}{\ensuremath{\mathrm{~Ge\kern -0.1em V\!/}c}}
\newcommand{\MeVc}{\ensuremath{\mathrm{~Me\kern -0.1em V\!/}c}}
\newcommand{\GeVcc}{\ensuremath{\mathrm{~Ge\kern -0.1em V\!/}c^2}}
\newcommand{\MeVcc}{\ensuremath{\mathrm{~Me\kern -0.1em V\!/}c^2}}
\newcommand{\MeV}{\ensuremath{\mathrm{~Me\kern -0.1em V}}}

\begin{document}

\title{The \boldmath $\boldmath X(3872)$ at the Te$\!$Vatron}


\author{G. Bauer\\
\small (Representing the CDF \& D\O\ Collaborations)
}{
  address={Laboratory of Nuclear Science, Massachusetts Institute of Technology, 
77 Massachusetts Avenue, Cambridge, MA 02139, USA
}}

\begin{abstract}
I report results on the $X(3872)$ from the Tevatron.
Mass and other properties have been studied,
with a focus on new results on the dipion mass
spectrum in $X\rightarrow J/\psi\,\pi^+\pi^-$ decays.
Dipions favor interpreting the decay as $J/\psi\,\rho$,
implying even $C$-parity for the $X$.
Modeling uncertainties do not allow distinguishing
between $S$- and $P$-wave decays of the $J/\psi$-$\rho$ mode.
Effects of $\rho$-$\omega$ interference in $X$ decay are
also introduced.
\end{abstract}

~\vspace*{-1.5cm}\\
{\small\raggedleft
 FERMILAB-CONF-05-488-E\\
PANIC`05~~~~~~~\\
(24-28 Oct 2005)
\vspace*{-0.4cm}

}

\maketitle

The charmonium-like $X(3872)$ stands as a major spectroscopic puzzle.
Its mass and what is known of its decays
makes $c\bar{c}$ assignments problematic.
Exotic interpretations have been offered,
notably that the $X$  may be a  $D^0$-$\overline{D}{^{*0}}$
``molecule''~\cite{XRev}.

\begin{figure}[b]
  \includegraphics[height=.23\textheight]{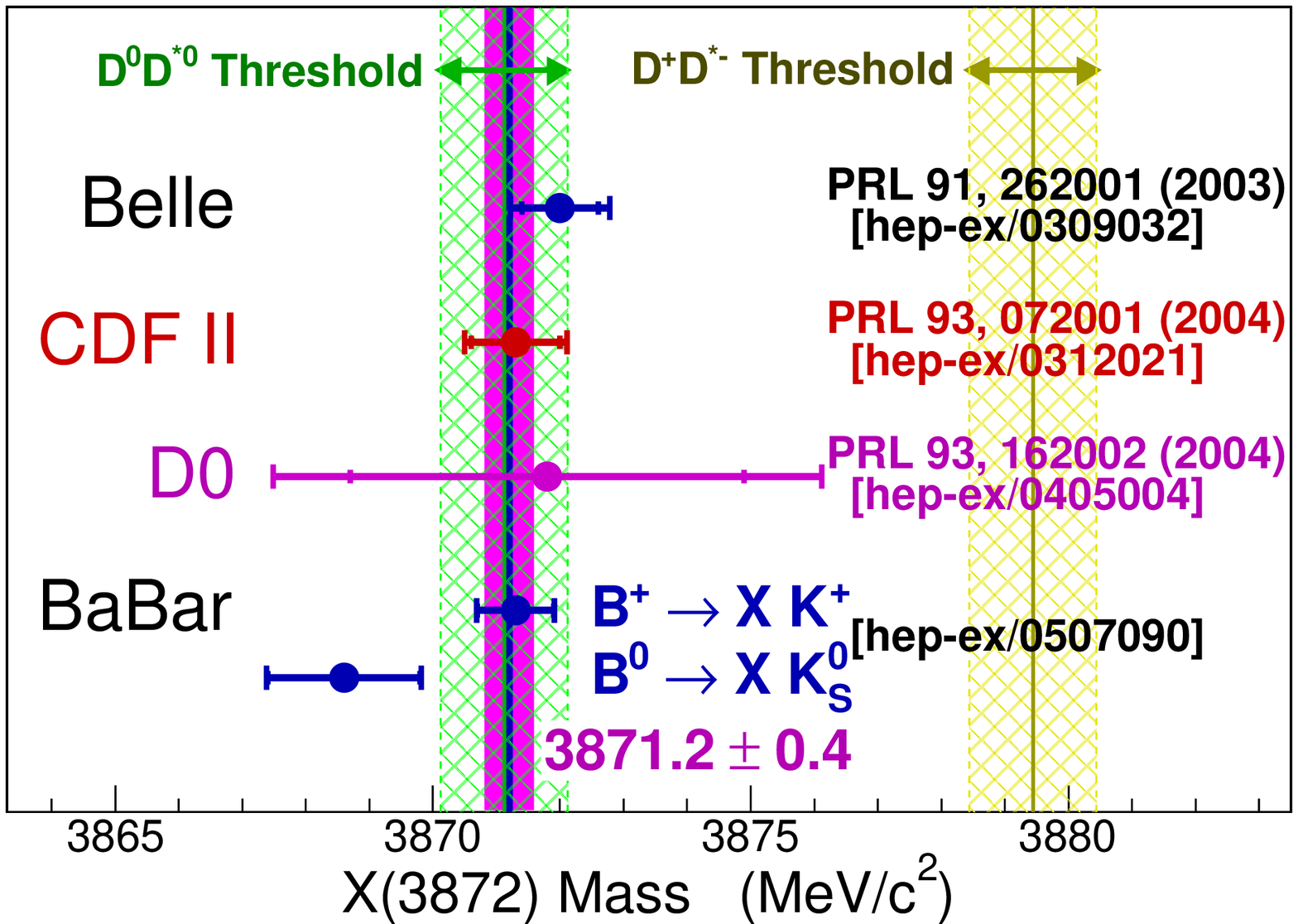} \includegraphics[height=.24\textheight]{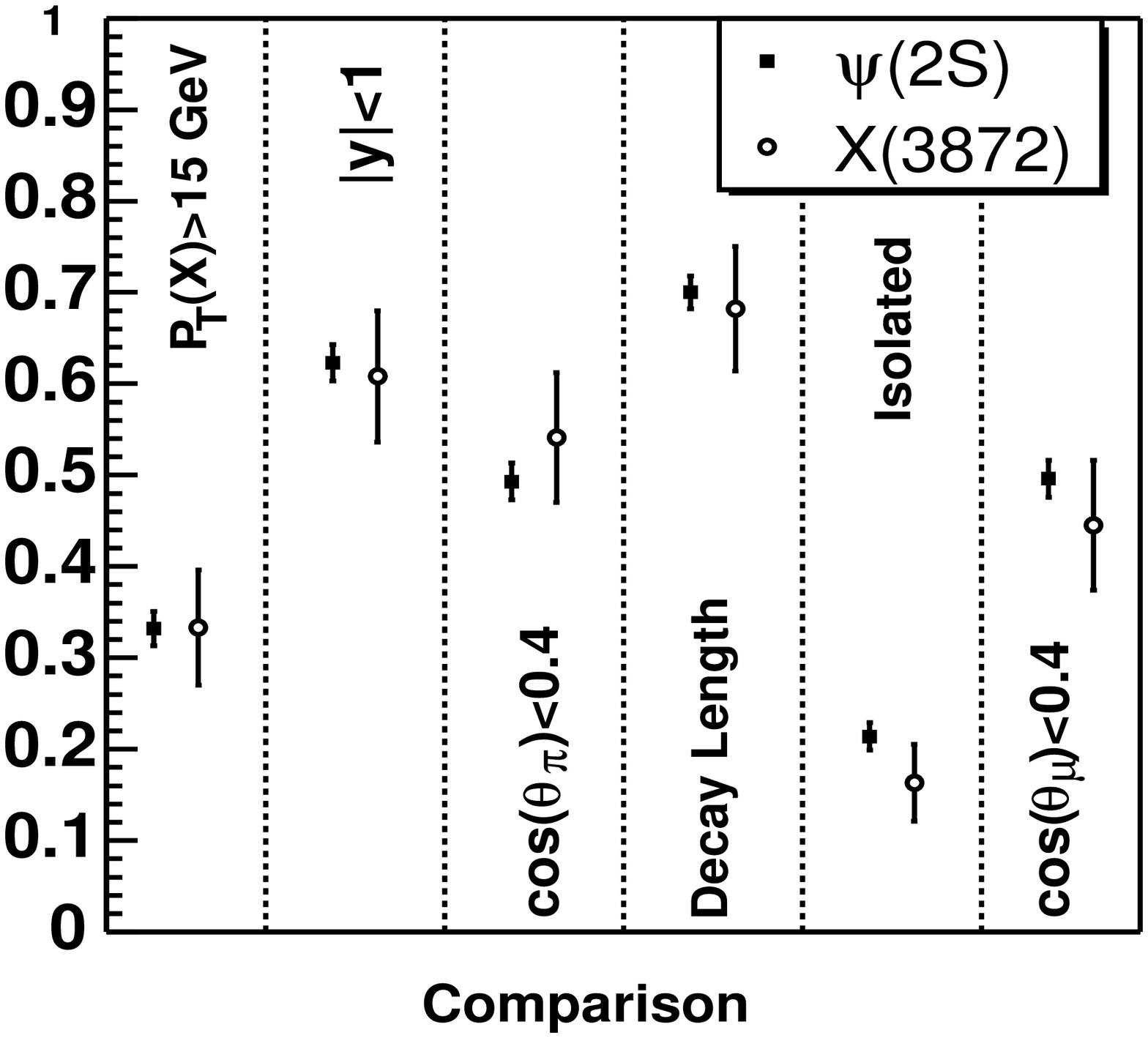}
\caption{{\bf LEFT:} Summary of $X$-mass measurements compared
          to the  $D^0\overline{D}{^{*0}}$ and $D^+{D}{^{*-}}$ thresholds.
  {\bf RIGHT:} D\O's comparison of $X$ production/decay properties to that of the $\psi(2S)$~\protect\cite{XD0}.
      The fraction of the yield surviving the listed  cut is plotted (see text for descriptions).
\label{Fig:Xmass}
}
\end{figure}

$X(3872)\rightarrow J/\psi\,\pi^+\pi^-$ was confirmed by  CDF~\cite{XCDF} and 
D\O~\cite{XD0}, and is  copiously produced at 
Fermilab's $\bar{p}p$ collider---albeit with high backgrounds.
Mass measurements are compared in Fig.~\ref{Fig:Xmass},
with an average of $3871.2\pm 0.4\MeVcc$.
D\O\  studied other  $X$ properties by comparing
the fractions of $X$ yield in various types of subsamples
to the corresponding fractions for the $\psi(2S)$~\cite{XD0}.
The results for  230 pb$^{-1}$  are summarized in Fig.~\ref{Fig:Xmass}, where the
subsamples are the fraction of signal which have:
{\bf a)}~$p_T(J/\psi\pi\pi)>15$ GeV,
{\bf b)}~$|y(J/\psi\pi\pi)|<1$, 
{\bf c)}~$\cos(\theta_\pi) < 0.4$ ($\pi$ helicity angle),
{\bf d)}~proper decay length $ct<100\,\mu$m,
{\bf e)}~no tracks with $\Delta R <0.5$ around the candidate,
{\bf f)}~$\cos(\theta_\mu) < 0.4$ ($\mu$~helicity angle).
In all cases the $X$ results are compatible
with those of the $\psi(2S)$.
CDF used the proper decay length $ct$ to quantify
the fraction of $X$'s that come from $b$-hadrons,
versus those that are promptly produced.
Using  220 pb$^{-1}$, CDF finds the fraction
of  $X$'s from $b$-decays is $16.1\pm\!4.9\pm\!2.0$\%,
in contrast to  $28.3\pm\!1.0\pm\!0.7$\% of $\psi(2S)$'s~\cite{XCDFLife}.
The $X$ fraction is somewhat lower than the $\psi(2S)$'s,
but within $\sim\!2\sigma$.
From these perspectives the $X$ is compatible with the $\psi(2S)$.
The large $X$-production  at the Tevatron is indicative
to some of a charmonium character~\cite{XProd}.
Na\"{\i}vely one expects  production of a fragile \mbox{$D^0$-$\overline{D}{^{*0}}$} molecule,
bound by an$\MeV$ or less, to be suppressed.
It may, however,  be sufficient to accommodate  these features if
the $X$ merely has  a significant $c\bar{c}$ ``core.''

Another property is the dipion mass spectrum.
If the $X$ has even $C$-parity, the dipions are (to lowest $L$) 
in a $1^{--}$ isovector state, and dominated by the $\rho^0$.
An odd-$C$ state produces $0^{++}$ dipions, for which
QCD multipole expansion predictions exist for $c\bar{c}$.

\begin{figure}[t]
\includegraphics[height=.23\textheight]{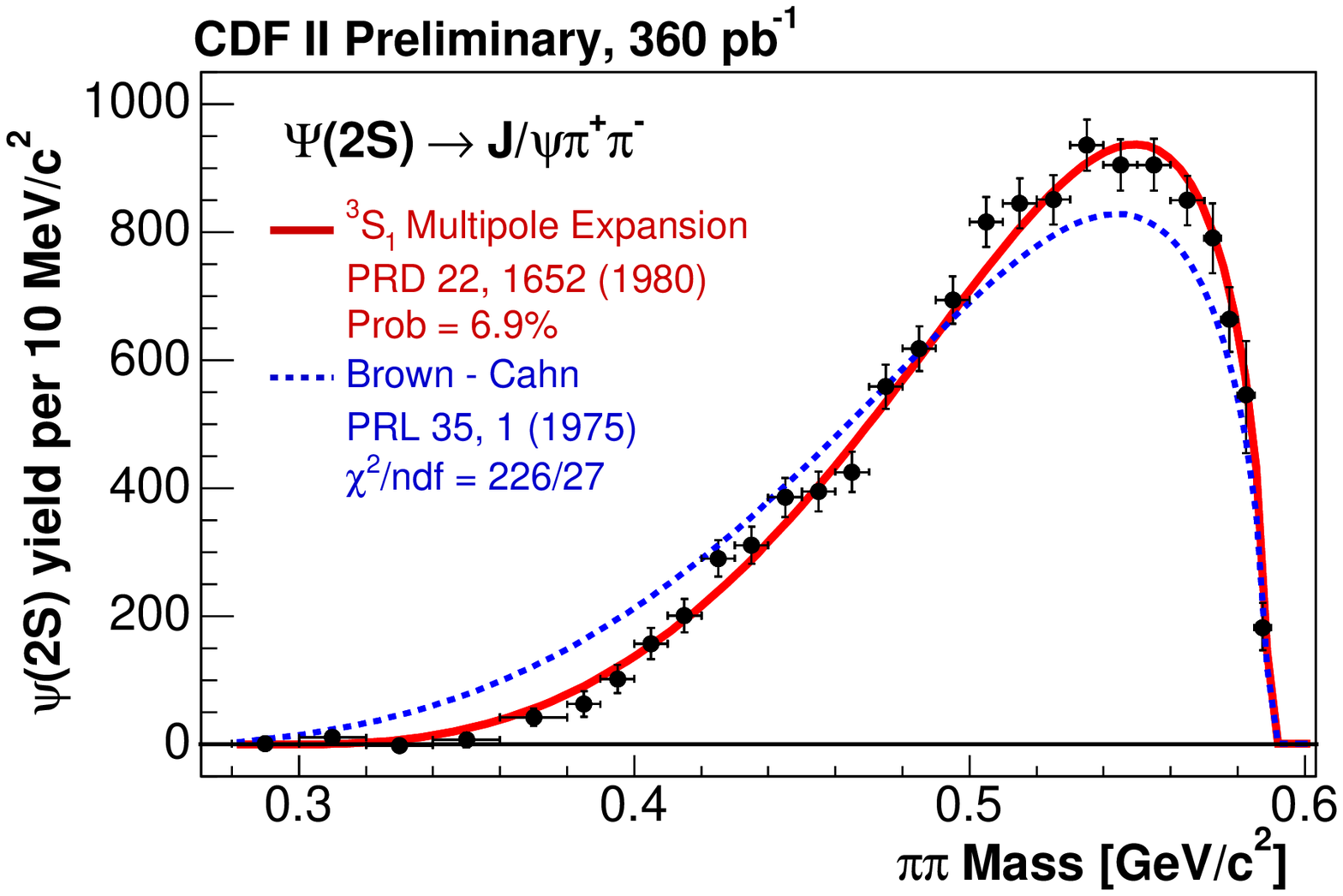}\includegraphics[height=.24\textheight]{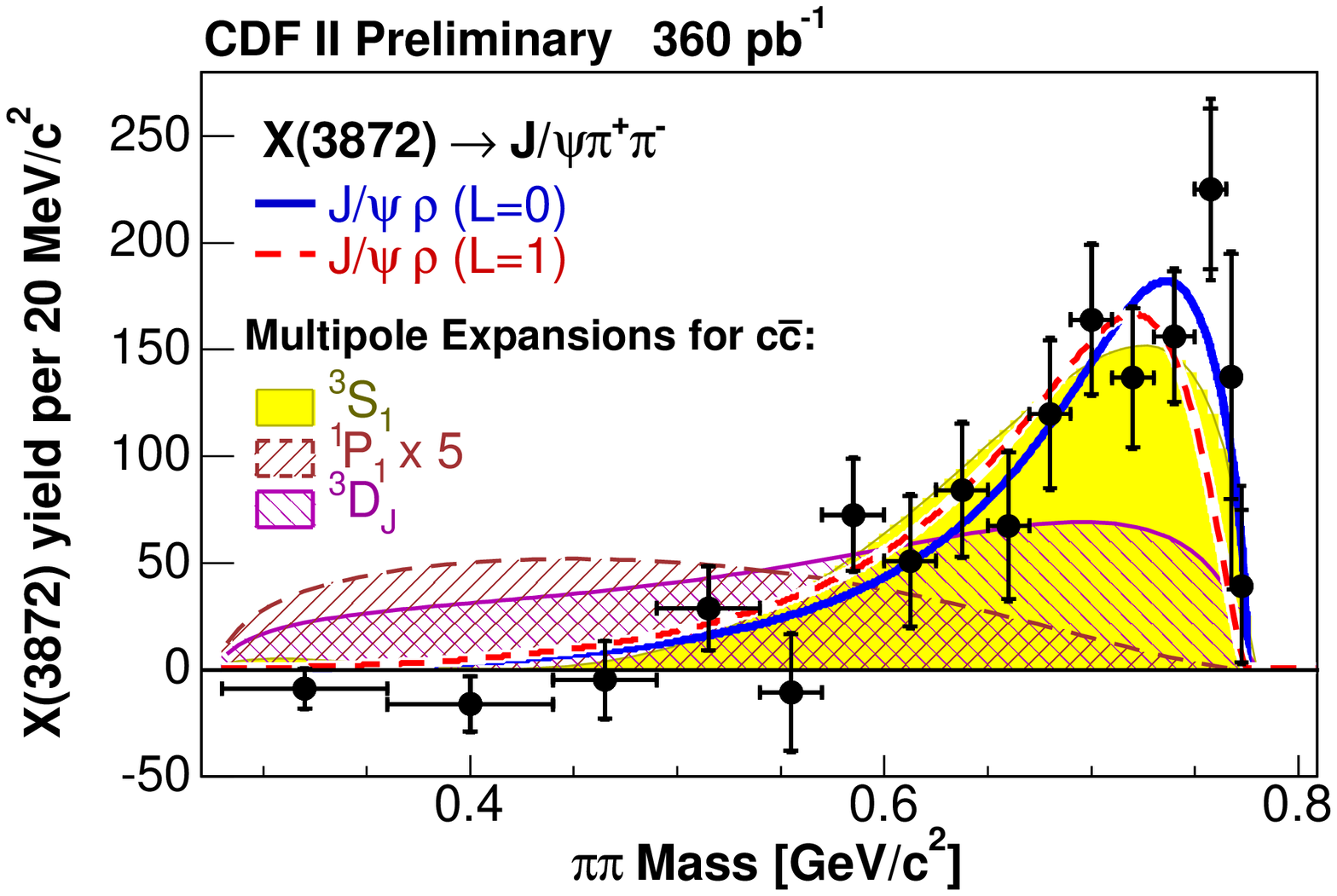}
\vspace*{8pt}
\caption{
  {\bf LEFT:} 
  The $\psi(2S)$ dipion mass spectrum with fits of $^3S_1$ 
   multipole expansion and an older calculation of Brown and Cahn.
  {\bf RIGHT:} 
  The $X(3872)$  dipion mass spectrum with fits
  of multipole expansion predictions for 
  $C$-odd charmonia, and of $X\rightarrow J/\psi\,\rho$
  for $L=0$ and 1 using a relativistic Breit-Wigner with
  Blatt-Weisskopf factors ($R_\rho=0.3$ and $R_X=1.0$ fm). 
 \label{Fig:XCDFpipi}
}
\end{figure}

CDF  used  360 pb$^{-1}$ ($\sim\!\!1.3$k $X$'s) 
to measure the $\pi\pi$-spectrum~\cite{MpipiCDFa,MpipiCDFb}.
The sample is divided
into $m_{\pi\pi}$ ``slices'' and fitted for $X(3872)$  and  $\psi(2S)$  yields.
After modest efficiency corrections, the spectra of Fig.~\ref{Fig:XCDFpipi}
were obtained.
The  $\psi(2S)$ is a good reference signal and is well  modeled
by  multipole predictions~\cite{Yan}.
Also  in  Fig.~\ref{Fig:XCDFpipi}
 are multipole fits to the $X$ for the $C$-odd $c\bar{c}$ states.
The $^1P_1$ and  $^3D_J$ fits are unacceptable.
The  $^3S_1$ is an excellent fit to the $X$,
but no  $^3S_1$ $c\bar{c}$ is available for assignment
in this mass region.

Earlier this spring CDF provided  $J/\psi\,\rho$ fits  using
a simple non-relativistic Breit-Wigner sculpted by phase space~\cite{MpipiCDFa}.
Good agreement was obtained (36\% probability).
About the same time Belle released new dipion data fit with a more sophisticated
$\rho$ model~\cite{Belle1PP}, which included the effects of
angular momentum $L$ in the $J/\psi$-$\rho$ system.
The phase-space factor of the $J/\psi$ momentum in the $X$ rest-frame, $k^*$,
generalizes to $(k^*)^{2L+1}$, thereby turning off the mass spectrum
at the upper kinematic limit ($k^*\rightarrow 0$) faster
for $L=1$ than for $L=0$.
Belle  obtained a good fit for $S$-wave decay, but only a $0.1\%$ probability
for  $L=1$.
Thus, the latter case was strongly disfavored, and in conjunction
with angular information, Belle argued for a $1^{++}$ assignment
for the $X$~\cite{Belle1PP}.
The above CDF fit for $J/\psi\,\rho$  was implicitly for $L=0$.
A CDF fit  using  Belle's  $L=1$ model also yields  $0.1\%$ probability.
The implication is, however, not robust.

Breit-Wigner formulations are often modified by  Blatt-Weisskopf form factors~\cite{BlattWeiss}.
The centrifugal modification to $(k^*)^{2L+1}$ 
tends to be too strong,
and for $L=1$ it is multiplied  by   $ f_{1i}(k^*) \propto (1 + R^2_i k^{*2})^{-\frac{1}{2}}$,
where $R_i$ is a  ``radius'' of  \mbox{meson-$i$}.
Specifically, CDF's  $J/\psi\,\rho$ model is:
$dN/dm_{\pi\pi}  \propto$ $(k^*)^{2{L}+1} f^2_{LX}(k^*) |B_\rho|^2$
for angular momentum $L$.
The $\rho$ propagator $B_\rho
\propto  \sqrt{ m_{\pi\pi}       
\Gamma_\rho(m_{\pi\pi})}/[  m_\rho^2 - m_{\pi\pi}^2   - i m_\rho\Gamma_\rho(m_{\pi\pi}) ]$, 
where
$\Gamma_\rho(m_{\pi\pi}) = 
\Gamma_0      \left[ q^*/q^*_0 \right]^3 \times
                     [m_\rho /m_{\pi\pi}]
                     [f_{1\rho}(q^*)/f_{1\rho}(q^*_0)]^2 $,
$q^*$ is the  $\pi$ momentum in the $\pi\pi$ rest-frame,
and $q^*_0 \equiv q^*(m_\rho)$.
The $\rho$ parameters $ m_\rho$ and $\Gamma_0$ are 
taken from the PDG.
The $L=0$ factor is  $f_{0i}(x) = 1$.
The  $f_{1i}$ factors require two uncertain parameters, $R_X$ and $R_\rho$. 
For light mesons, like the $\rho$, values $\sim\!0.3$~fm are usually found,
whereas for charm mesons larger radii $\sim\!1$~fm are often used~\cite{Rstuff}.
Choosing these values for $R_\rho$ and $R_X$, CDF obtains the fits
in Fig.~\ref{Fig:XCDFpipi} (Right).
The $L=0$ fit has an excellent probability  of 55\%.
While the $L=1$ probability is not quantitatively as good, 
it is a respectable 7.7\%.
This $P$-wave fit is sensitive to the $R_i$'s,
whereby the probability  can be increased by lowering 
$R_\rho$ and/or raising  $R_X$.
We conclude that  flexibility in the
fit model can  accommodate either $L$.

Other modeling uncertainties may  arise, for example,
the effects of $\rho$-$\omega$ interference.
Belle reported  $X\rightarrow J/\psi\, \pi^+\pi^-\pi^0$,
and interprets it as decay via a virtual $\omega$.
As such, they find the ratio of 
$ J/\psi\,\omega$ to $ J/\psi\,\rho$
branching ratios ${\cal R}_{3/2}$
is $1.0 \pm 0.5$~\cite{BellePsiGammaOmega}.
Although $\omega \rightarrow \pi^+\pi^-$ is  nominally negligible here, 
its interference effects may not be.

$dN_{2\pi}/dm_{\pi\pi}$ is generalized
by replacing  $|B_\rho|^2$ with $|A_\rho B_\rho + e^{i\phi}A_\omega B_{\omega2\pi}|^2$
where $A_\rho$ and $A_\omega$ are  $X$-decay amplitudes
via $\rho$ and $\omega$, and $\phi$ is the relative phase.
The form for  $B_{\omega2\pi}$ is identical to $B_{\rho}$ except
$\rho$ quantities are replaced  by $\omega$ ones,
including 
the  $\omega\rightarrow \pi\pi$ branching ratio.
The ratio  $|A_\omega/A_\rho|$ is
established 
by the relationship
between  ${\cal R}_{3/2}$ and the integrals of  $dN_{2\pi}/dm_{\pi\pi}$ 
and  $dN_{3\pi}/dm_{3\pi}$ for $X\rightarrow J/\psi\,\pi^+\pi^-\pi^0$,
where the latter is $\propto  |A_\omega B_{\omega3\pi}|^2$. 
The $B_{\omega3\pi}$ follows  $B_{\omega2\pi}$ except
the numerator contains $\Gamma_{\omega3\pi}(m)$.
While $\Gamma_{\omega2\pi}(m)$ 
follows $\Gamma_\rho(m)$,
a different form for $\Gamma_{\omega3\pi}(m)$
is adapted from the SND experiment studying $e^+e^-\!\rightarrow\!\pi^+\pi^-\pi^0$~\cite{Achasov}.
They model $\omega\rightarrow \pi^+\pi^-\pi^0$   
as virtual $\rho\pi$ decays and use the 
$\omega$ matrix-element
$|\vec{q}_{\pi^+} {\boldmath \times} \vec{q}_{\pi^-}  |^2$,
where $\vec{q}_{\pi^{+/-}}$ are  $\pi^{+/-}$  momenta.

\begin{figure}[t]
  \includegraphics[height=.23\textheight]{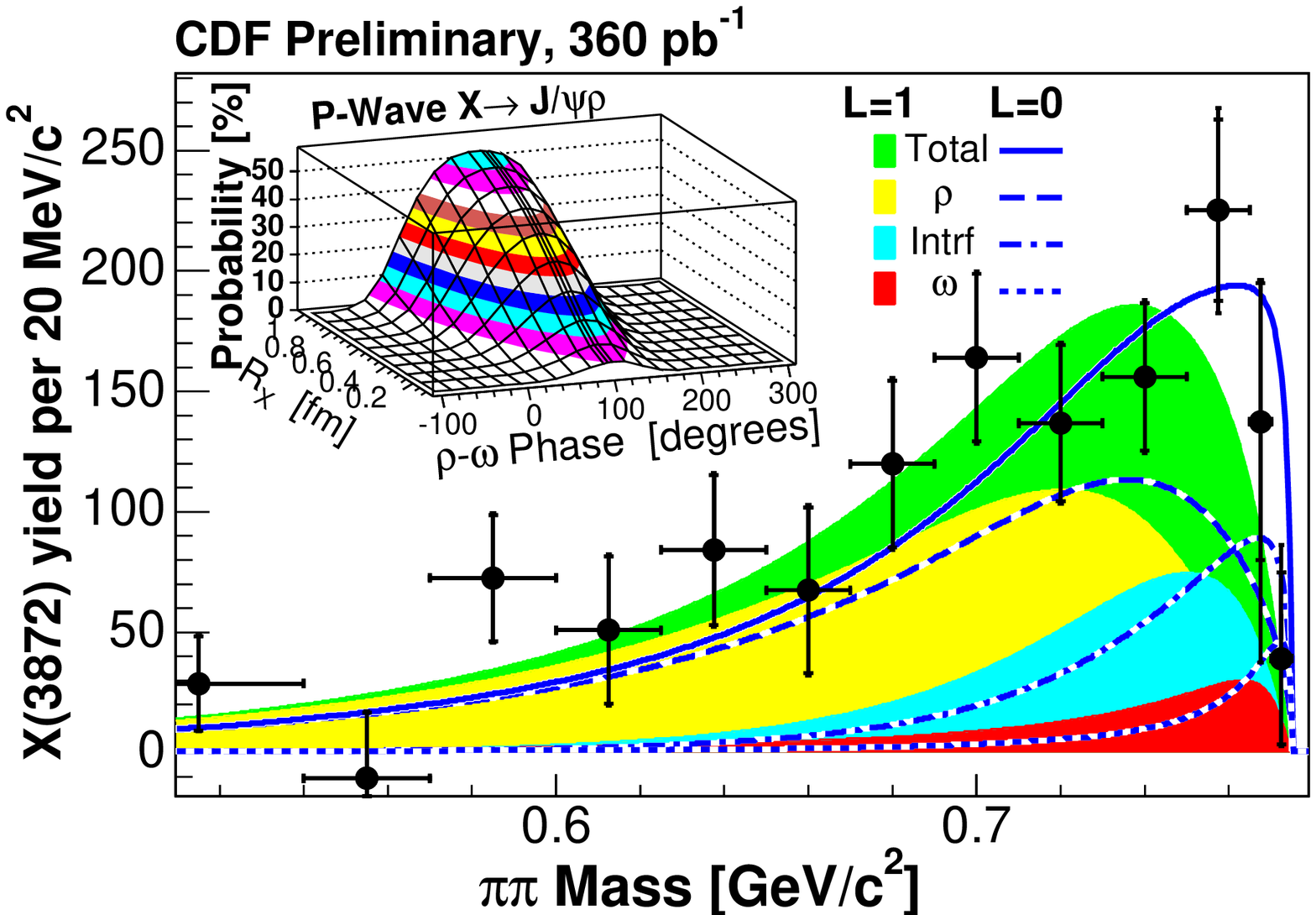}\includegraphics[height=.21\textheight]{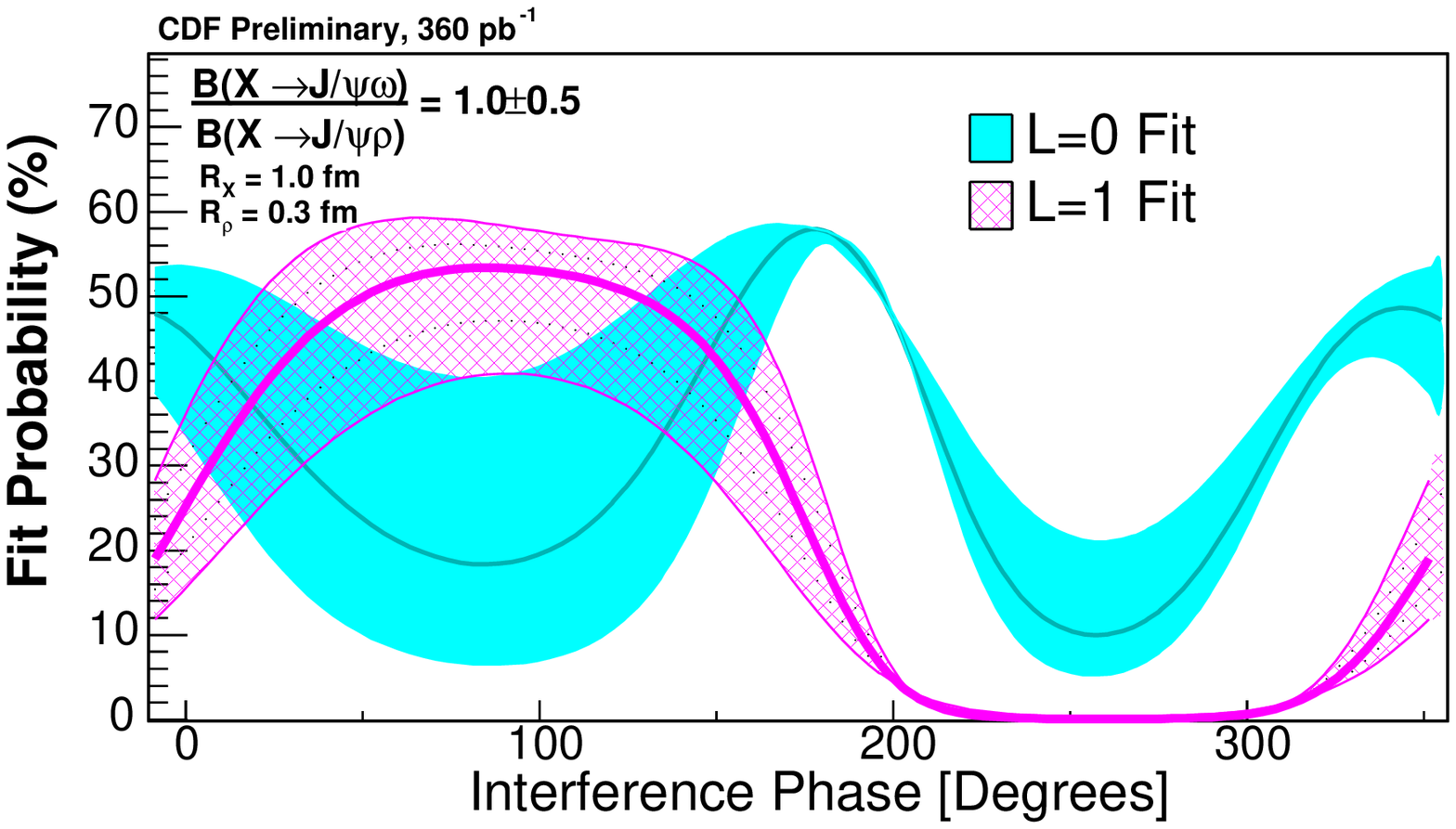}
\vspace*{8pt}
\caption{   
 {\bf LEFT:} 
 Blow-up of the dipion  spectrum with $J/\psi\,\rho$ fits 
 for $L=0$ (lines) and 1 (shaded)  including
 $\rho$-$\omega$ interference with $95^\circ$ phase and 
 sub-components  set by  ${\cal R}_{3/2}=1.0$.
 The decomposition into  $\rho$, interference, and $\omega$ terms
 is given.
 The inset shows $L=1$ fit probabilities as a function
 of $\phi$ and $R_X$ in 5\% contours.
 {\bf RIGHT:} 
 $J/\psi\,\rho$ fit probabilities for  $L=0$ (shaded) and 1 (hatched) 
 as a function of phase. The bands span the $\pm1\sigma$ range of   ${\cal R}_{3/2}$.
\label{Fig:XCDFpipi2}
}
\end{figure}

The integral  of  $dN_{2\pi}/dm_{\pi\pi}$  depends upon the phase, 
which is {\it a priori} unknown.
As an illustration, $|A_\omega/A_\rho|$ is  determined
assuming that  $\phi$  arises completely
from $\rho$-$\omega$ mixing, i.e. $\phi =95^\circ$~\cite{ExpectPhi}.
The  $dN_{2\pi}/dm_{2\pi}$  decomposes into three parts:
``pure''  $\rho$ and $\omega$ terms, and an interference cross-term.
In this model with  ${\cal R}_{3/2}=1.0$,
these  fractions are, 
respectively,  71.0, 6.2, and 22.8\% for  $S$-wave decay,
and  67.4, 8.7, 23.9\% for  $P$-wave.
Fits with these fractions imposed
are shown in Fig.~\ref{Fig:XCDFpipi2} (Left).
The $S$-wave probability has  declined as the model
peaks too much at high mass, but is still very good at 19\%.
Increasing the amount of  high masses with interference 
improves the $P$-wave fit to 53\%.
The $L=1$ fit is  sensitive to $\phi$
and $R_X$ as is seen in the inset of  Fig.~\ref{Fig:XCDFpipi2}.
The dependence on $R_\rho$ is relatively weak for both $L$.
The overall picture from these fits is insensitive
to the $\pm1\sigma$ span of  ${\cal R}_{3/2}$,
as is seen in  Fig.~\ref{Fig:XCDFpipi2} (Right).

In summary, properties of  $X(3872)\rightarrow J/\psi\,\pi^+\pi^-$ 
studied at the Tevatron are quite similar to those of the $\psi(2S)$.
There is no viable $C$-odd charmonium  
assignment according to QCD multipole expansion fits to the $\pi\pi$-mass spectrum.
Decay to $J/\psi\,\rho$ provides good fits,
irrespective of the $c\bar{c}$ structure.   
This implies the $X$ is $C$-even, in-line with Belle's report
of $X\rightarrow J/\psi\,\gamma$~\cite{BellePsiGammaOmega}.
The effects of  $\rho$-$\omega$ interference are introduced,
and can be quite important.
This type of  $\rho$-$\omega$ modeling highlights 
  that ${\cal R}_{3/2}  \sim 1$ implies
  the {\it intrinsic} amplitude for  $X\rightarrow J/\psi\,\rho$
  is actually significantly suppressed relative to  $J/\psi\,\omega$ 
  by virtue of the much greater
  phase space for $J/\psi\,\rho$ decay over  $J/\psi\,\omega$.
Given the modeling uncertainties governing the tails of the Breit-Wigners---especially
if {$\rho$-$\omega$} interference is in play---the CDF spectrum can be well 
described by  $J/\psi\,\rho$ decay of either $L=0$ or 1: such as
from $C$-even charmonia (e.g.~$1^{++}$ or  $2^{-+}$) or
by a $1^{++}$ exotic as preferred for a $D^0$-$\overline{D}{^{*0}}$ molecule.


\end{document}